\documentclass[10pt,twocolumn,twoside]{IEEEtran} 

\usepackage{amsmath}        
\usepackage{amssymb}        
\usepackage{amsthm}         

\theoremstyle{plain}
\newtheorem{lemma}{Lemma}
\theoremstyle{definition}
\newtheorem{definition}{Definition}
\newtheorem{problem}{Problem}
\theoremstyle{remark}
\newtheorem{remark}{Remark}

\usepackage[numbers,sort&compress]{natbib}   

\usepackage{enumitem}       
\usepackage{graphicx}
\usepackage{float}          
\usepackage{algorithm}
\usepackage{algpseudocode}  
\usepackage{comment}        

\usepackage{tikz}
\usetikzlibrary{positioning, arrows.meta}

\usepackage{url}
\usepackage[hidelinks]{hyperref}             

\newcommand{\stateVector}{\mathbf{x}}
\newcommand{\inputVector}{\mathbf{u}}

\newcommand{\ellipsoidalSet}{\mathcal{E}}    
\newcommand{\levelSet}{\mathcal{B}}               
\newcommand{\funnel}{\mathcal{F}}

\newcommand{\subroutine}[1]{\textsc{#1}}     
  {\par\centering\begin{minipage}{#1}}%
  {\end{minipage}\par}

\providecommand{\figref}[1]{Fig.~\ref{#1}}

\begin{document}

\title{Computing Invariant Reachable Sets through\\Sum-of-Squares Programming}

\author{
\IEEEauthorblockN{Mohamed Khalid~M Jaffar}
\thanks{\indent *The author is with the Department of Aerospace Engineering, University of Maryland, College Park, USA. Email: {\tt\footnotesize khalid26@umd.edu}.
This work was not supported by any organization.}
}

\markboth{Journal of \LaTeX\ Class Files,~Vol.~14, No.~8, August~2021}%
{M~Jaffar: Computing invariant reachable sets using SOS programming}


\maketitle

\begin{abstract}
Funnels, time-varying invariant sets around nominal trajectories, certify that a closed-loop system tracks its plan, and can be composed into verified motion plans. Their computation by sum-of-squares (SOS) programming is well established in formulation but not in practice: the result depends on implementation choices such as the initial guess, Lyapunov function normalization, and termination rule, which the literature documents only briefly. This paper presents a self-contained treatment of SOS-based funnel computation, implemented in the open-source package \texttt{computeSOSFunnels}. We propose a minimum-volume formulation with inlet and outlet constraints under which funnels with coinciding nominal endpoints compose by construction. Numerical experiments on a unicycle, with timing studies on a cart-pole and a 12-state quadrotor, characterize the pipeline: infeasible initial guesses are partly decidable before solving, the certificate shows no violations among 39,000 boundary samples per funnel under the true dynamics, and constructing the semidefinite program, not solving it, dominates computation time. 
\end{abstract}


\section{Introduction}
\label{sec:intro_funnel_chapter}

Autonomous systems that act near people and obstacles need more than a nominal motion plan: they need a guarantee that the closed-loop system stays near that plan despite deviations from it. Funnels provide such a guarantee. A funnel is a time-varying set around a nominal trajectory that every closed-loop trajectory starting in its inlet remains inside until it exits through its outlet. Because the guarantee is set-based, when funnels compose, that is, when the outlet of one funnel lies in the inlet of the next, the guarantees chain, and a sequence of funnels becomes a verified motion plan \citep{burridge1999sequential}. This composability makes funnels a natural primitive for feedback motion planning \citep{Tedrake.etal.IJRR10, Majumdar.Tedrake.IJRR17, Jaffar.Otte.WAFR22}.

Computing funnels for a general dynamical system is the bottleneck. For nonlinear systems, a standard approach is to search for time-varying Lyapunov functions with sum-of-squares (SOS) programming, which turns invariance conditions into semidefinite programs (SDPs) \citep{parrilo2000structured, papachristodoulou2002construction, Tobenkin.etal.IFAC11}. The resulting problem is bilinear (non-convex) in the Lyapunov function and the SOS multipliers, so it is solved by alternating between convex subproblems, and its solution depends on design choices the formulation leaves open --- the initial guess, termination rule, normalization that fixes the scale of the Lyapunov function, and the polynomial approximation of the dynamics. These choices decide whether the method returns a useful funnel or an infeasible program, yet the literature typically documents them only briefly, and the theory, which draws on real algebraic geometry and semidefinite optimization is spread across several papers. This paper is written to be self-contained, so that a reader new to SOS programming can follow it from theory to formulation to implementation.

\textit{SOS funnels and their use in planning:} \citet{Tedrake.etal.IJRR10} presented LQR-trees that grow a tree of trajectories, each stabilized by a time-varying linear quadratic regulator (TVLQR), and verify a funnel around each with SOS programming. \citet{Tobenkin.etal.IFAC11} developed the underlying funnel computation framework, including the time-sampled verification and the Riccati-based initial guess adopted in this work. \citet{Majumdar.Ahmadi.Tedrake.ICRA13} extended the search to the controller itself, maximizing the set of states driven into a goal region. \citet{Majumdar.Tedrake.IJRR17} reversed the objective, computing minimum-volume outer approximations of the states reachable from a given initial set; our formulation is closest to theirs. Funnels have since served as primitives for online replanning in dynamic environments \citep{jaffar2022pip}. 

\textit{Alternatives to SOS-based funnels:} Funnel synthesis for systems with Lipschitz nonlinearities can be posed as a numerical optimal control problem with matrix inequality constraints~\citep{Kim2023}. Hamilton--Jacobi reachability analysis yields tracking-error bounds for planning with reduced-order models \citep{Herbert.etal.CDC17}, and Bernstein polynomial bounds give fast funnel computation for polynomial systems \citep{Seo2021}. Contraction metrics certify tracking tubes through differential conditions rather than Lyapunov sublevel sets \citep{Singh2023}. These methods trade the generality of polynomial certificates for scalability or tighter structure. SOS methods face their own scalability limit, since the SDP size grows combinatorially with state dimension and polynomial degree; diagonally dominant relaxations \citep{ahmadi2019dsos} and related techniques \citep{majumdar2020recent} trade conservatism for tractability.

\textit{Gap:} The SOS formulation for computing funnels is well established, but its practical implementation is not. Where reference implementations exist, the accompanying papers do not fully characterize how the computation behaves: which initial guesses yield feasible programs, when to stop the alternation procedure, how well a polynomial approximate certificate holds for the true dynamics, and where the computation time goes. This paper addresses that gap. Its contributions are:
\begin{enumerate}
    \item \texttt{computeSOSFunnels}\footnote{\href{https://github.com/khalid2696/computeSOSFunnels}{\texttt{github.com/khalid2696/computeSOSFunnels}}}, an open-source, end-to-end implementation of the computation pipeline, from trajectory optimization and TVLQR synthesis to polynomial approximation and invariance certification. It automatically generates Taylor approximations of arbitrary user-specified degree from symbolic system dynamics. 
    \item An empirical characterization of the pipeline in practice: feasibility of initial guesses, part of which is decidable before any SDP is solved; validity of the certificate under the true dynamics; and computational cost, including a closed-form count of the SDP size.
    \item A minimum-volume funnel formulation with both inlet and outlet constraints. As in \citep{Majumdar.Tedrake.IJRR17}, the funnel contains a given inlet set; in addition, its outlet lies within a user-specified terminal set nested inside the inlet set. So funnels with coinciding nominal endpoints compose, avoiding runtime set-containment checks whose cost accumulates during motion planning.
\end{enumerate}


\section{Preliminaries and Problem Setting}
\label{sec:funnel_preliminaries}

\textit{Notation and Terminology:}
$\mathbb{S}^n$ denotes the set of $n \times n$ real symmetric matrices and $\mathbb{S}^n_{++} \subset \mathbb{S}^n$ is its positive definite subset. $\mathbf{A} \succeq \mathbf{B}$ ($\mathbf{A} \succ \mathbf{B}$) means $\mathbf{A} - \mathbf{B}$ is positive semidefinite (definite). $\mathbf{1}_n$ is the vector of ones in $\mathbb{R}^n$. For $\mathbf{c} \in \mathbb{R}^n$ and $\mathcal{S} \subseteq \mathbb{R}^n$, $\mathbf{c} + \mathcal{S} \triangleq \{\mathbf{c} + \mathbf{s} : \mathbf{s} \in \mathcal{S}\}$, and $\mathrm{vol}(\mathcal{S})$ is the Lebesgue volume of $\mathcal{S}$. 
$\mathbb{R}[\mathbf{z}]$ is the ring of polynomials in $\mathbf{z} \in \mathbb{R}^n$ with real coefficients, $\mathbb{R}[\mathbf{z}]_d$ is its subset of polynomials of degree at most $d$, and $\Sigma[\mathbf{z}] \subset \mathbb{R}[\mathbf{z}]$ is the cone of sum-of-squares (SOS) polynomials (Section~\ref{subsec:funnel_sos}).
For $\mathbf{M}_\star \in \mathbb{S}^n_{++}$, $\ellipsoidalSet_\star \triangleq \{\mathbf{z} : \mathbf{z}^\top \mathbf{M}_\star \mathbf{z} \le 1\}$ is a solid ellipsoid centered at the origin with shape matrix $\mathbf{M}_\star$. For two such ellipsoids, $\ellipsoidalSet_2 \subseteq \ellipsoidalSet_1$ if and only if $\mathbf{M}_2 \succeq \mathbf{M}_1$.

Consider a system with dynamics given by
\begin{equation}
    \dot{\stateVector} = \mathbf{f}(\stateVector, \inputVector), \quad \stateVector \in \mathbb{R}^n, \; \inputVector \in \mathcal{U}_{\mathrm{feas}} \subseteq \mathbb{R}^m,
    \label{eq:nonlinear_system_dynamics_funnel_chapter}
\end{equation}
where $\mathbf{f}$ is at least three times continuously differentiable and $\mathcal{U}_{\mathrm{feas}}$ encodes actuator limits. Given a nominal state-input pair $(\stateVector_{nom}(t), \inputVector_{nom}(t))$ satisfying \eqref{eq:nonlinear_system_dynamics_funnel_chapter} at all $t \in [t_0, t_f]$ and a tracking controller, the deviation $\bar{\stateVector} \triangleq \stateVector - \stateVector_{nom}$ evolves under closed-loop deviation dynamics $\dot{\bar{\stateVector}} = \bar{\mathbf{f}}(\bar{\stateVector}, t)$ with $\bar{\mathbf{f}}(\mathbf{0}, t) = \mathbf{0}$. Section~\ref{sec:funnel_computation_approach} constructs $\bar{\mathbf{f}}$ explicitly.

\begin{definition}[\textsc{Funnel} \citep{jaffar2022pip}, Defn.~1]
\label{def:funnel_introduction}
For closed-loop dynamics on $[t_0, t_f]$, a set $\funnel \subseteq [t_0, t_f] \times \mathbb{R}^n$ is a \textit{funnel} if, for every $(\tau, \stateVector_\tau) \in \funnel$, the solution with $\stateVector(\tau) = \stateVector_\tau$ satisfies $(t, \stateVector(t)) \in \funnel$ for all $t \in [\tau, t_f]$. Its cross-section at time $t$ is $\funnel(t) \triangleq \{\stateVector : (t, \stateVector) \in \funnel\}$. We call $\funnel(t_0)$ as the \textit{inlet} and $\funnel(t_f)$ the \textit{outlet}.
\end{definition}

Funnels become planning primitives through sequential composition \citep{burridge1999sequential}. Funnels $\funnel^a$ and $\funnel^b$ are \textit{sequentially composable} if
\begin{equation}
    \funnel^a(t^a_f) \subseteq \funnel^b(t^b_0),
    \label{eq:composability_stricter}
\end{equation}
so that system state gets handed off from the first funnel $\funnel^a$ to the subsequent $\funnel^b$ without leaving either. Condition \eqref{eq:composability_stricter} is stronger than planning requires. Let us partition the state into cyclic and non-cyclic coordinates, $\stateVector = (\stateVector_c, \stateVector_{nc})$, where the dynamics do not depend on $\stateVector_c$ (e.g., the planar position of a unicycle). A funnel shifted along $\stateVector_c$ is then again a funnel, because its closed-loop deviation dynamics are unaffected by the shift \citep{Majumdar.Tedrake.IJRR17}.

\begin{definition}[\textsc{Motion-plan composability} \citep{jaffar2022pip}, Defn.~2]
\label{def:motion_plan_composability_introduction}
Funnel pair ($\funnel^a$, $\funnel^b$) is \textit{motion-plan composable} if
\begin{equation}
    \mathcal{P}_{nc}\big(\funnel^a(t^a_f)\big) \subseteq \mathcal{P}_{nc}\big(\funnel^b(t^b_0)\big),
    \label{eq:runtime_composability}
\end{equation}
where $\mathcal{P}_{nc}(\cdot)$ is the orthogonal projection onto the non-cyclic coordinates.
\end{definition}

Under \eqref{eq:runtime_composability}, every state in the outlet of $\funnel^a$ lies in the inlet of some shift of $\funnel^b$ along $\stateVector_c$. At planning run time, $\funnel^b$ is shifted so that its inlet contains the actual hand-off state \citep{Jaffar.Otte.WAFR22}. Since projection preserves containment, \eqref{eq:composability_stricter} implies \eqref{eq:runtime_composability}.

\subsection{Semi-algebraic sets and polynomial approximation}
\label{subsec:funnel_semialgebraic}

A basic (closed) semi-algebraic set is a set of the form
\begin{equation}
    \mathcal{G} = \{\mathbf{z} \in \mathbb{R}^n : g_1(\mathbf{z}) \ge 0, \dots, g_r(\mathbf{z}) \ge 0\}, \quad g_i \in \mathbb{R}[\mathbf{z}].
    \label{eq:semialgebraic_set}
\end{equation}
Every set in this paper is of this form. For instance, ellipsoids $\ellipsoidalSet_\star$ use $g = 1 - \mathbf{z}^\top \mathbf{M}_\star \mathbf{z}$ and sublevel sets of a time-varying Lyapunov function, $\{\bar{\stateVector} : V(\bar{\stateVector}, t) \le \rho(t)\}$ use $g = \rho(t) - V(\bar{\stateVector}, t)$ for polynomial $V$, $\rho$. Certifying properties of such sets reduces to certifying nonnegativity of polynomials on them, which is the task SOS programming addresses \citep{blekherman2012semidefinite}.

This reduction requires polynomial functions. Robot dynamics are rarely polynomial; for example, a simple unicycle model itself contains $\cos$ and $\sin$ terms. Two remedies exist: 
(i) \textit{Recasting} introduces auxiliary states and equality constraints that represent non-polynomial terms exactly \citep{papachristodoulou2005analysis}, at the cost of a larger state dimension. 
(ii) \textit{Taylor approximation} replaces $\bar{\mathbf{f}}$ with a truncated expansion about the nominal trajectory \citep{Tobenkin.etal.IFAC11, Majumdar.Tedrake.IJRR17}; it preserves the state dimension but certifies the approximate dynamics only. In this work we adopt Taylor approximation and empirically verify the obtained certificates \textit{a posteriori} against the true dynamics (Section~\ref{subsec:results_discussion_funnel_chapter}).

\subsection{SOS optimization framework}
\label{subsec:funnel_sos}

\begin{definition}[\textsc{SOS polynomial}]
A polynomial $p \in \mathbb{R}[\mathbf{z}]_{2d}$ is a \textit{sum of squares} if $p = \!\sum_{i=1}^{r} q_i^2$ for some $q_1, \dots, q_r \!\in\! \mathbb{R}[\mathbf{z}]_d$.
\end{definition}

Every SOS polynomial is nonnegative, but the converse is not true in general. By Hilbert's classical result, nonnegativity and SOS coincide only for univariate polynomials, quadratics, and bivariate quartics. The Motzkin polynomial $z_1^4 z_2^2 + z_1^2 z_2^4 - 3 z_1^2 z_2^2 + 1$ is a counterexample: it is nonnegative but not SOS \citep{blekherman2012semidefinite}. This restriction is the price for tractability: checking nonnegativity is NP-hard already for fourth-degree polynomials, whereas SOS-based analysis reduces it to a semidefinite feasibility problem, which is solvable to any fixed accuracy in polynomial time. 

Let $\mathbf{m}_d(\mathbf{z})$ be the vector of all monomials in $\mathbf{z} \in \mathbb{R}^n$ of degree at most $d$; its size is $s = \binom{n+d}{d}$. Then $p \in \Sigma[\mathbf{z}]$ with $\deg p = 2d$ if and only if there exists a \textit{Gram matrix} $\mathbf{G} \in \mathbb{S}^{s}$ such that
\begin{equation}
    p(\mathbf{z}) = \mathbf{m}_d(\mathbf{z})^\top \mathbf{G}\, \mathbf{m}_d(\mathbf{z}), \qquad \mathbf{G} \succeq 0
    \label{eq:gram_matrix}
\end{equation}
\citep{parrilo2000structured, parrilo2003semidefinite}. Matching coefficients of $p$ on both sides of \eqref{eq:gram_matrix} yields linear equality constraints on the entries of $\mathbf{G}$. Consequently, an \textit{SOS program}, which minimizes a linear function of the decision variables subject to the constraint that polynomials with coefficients affine in those variables are SOS, is a semidefinite program (SDP) solvable by interior-point methods. Searching for Lyapunov functions this way originated from the seminal works of \citet{parrilo2000structured} and \citet{papachristodoulou2002construction}.

The Gram matrix size $\binom{n+d}{d} \times \binom{n+d}{d}$ drives computation cost and grows combinatorially in state dimension $n$ and the degree $d$. This is the principal obstacle to scaling SOS methods. It motivates restricting $\mathbf{G}$ to diagonally dominant or scaled diagonally dominant matrices (DSOS/SDSOS), which replaces the SDP with a linear or second-order cone program, respectively \citep{ahmadi2019dsos, majumdar2020recent}.

\subsection{Polynomial S-procedure}
\label{subsec:funnel_s_procedure}

SOS constraints certify global nonnegativity. Funnel invariance conditions instead require nonnegativity on sets or implications between polynomial inequalities. The polynomial S-procedure, a generalization of the classical S-procedure \citep{boyd1994linear}, converts both into SOS constraints using polynomial multipliers \citep{parrilo2003semidefinite}.

\begin{lemma}[\textsc{Polynomial S-procedure}]
\label{lem:s_procedure}
Let $p, g \in \mathbb{R}[\mathbf{z}]$.
\begin{enumerate}[label=(\alph*), itemsep=-0.25em]
    \item If there exists $\mu \in \Sigma[\mathbf{z}]$ such that $p - \mu\, g \in \Sigma[\mathbf{z}]$, then $\{\mathbf{z} : g(\mathbf{z}) \ge 0\} \subseteq \{\mathbf{z} : p(\mathbf{z}) \ge 0\}$, equivalently $g(\mathbf{z}) \ge 0\ \implies p(\mathbf{z}) \ge 0$ for all $\mathbf{z}$.
    \item If there exists $\lambda \in \mathbb{R}[\mathbf{z}]$ such that $p - \lambda\, g \in \Sigma[\mathbf{z}]$, then $\{\mathbf{z} : g(\mathbf{z}) = 0\} \subseteq \{\mathbf{z} : p(\mathbf{z}) \ge 0\}$, equivalently $g(\mathbf{z}) = 0\ \implies p(\mathbf{z}) \ge 0$ for all $\mathbf{z}$.
\end{enumerate}
\end{lemma}
\begin{proof}
(a) If $g(\mathbf{z}) \ge 0$, then $p(\mathbf{z}) \ge \mu(\mathbf{z})\, g(\mathbf{z}) \ge 0$, since $\mu(\mathbf{z}) \ge 0$. (b) If $g(\mathbf{z}) = 0$, then $p(\mathbf{z}) = \big(p - \lambda g\big)(\mathbf{z}) \ge 0$; no sign condition on $\lambda$ is needed because the multiplier term vanishes.
\end{proof}

Statement (a) certifies an implication on $\{g \ge 0\}$ (a set containment), and statement (b) certifies an implication on a level set $\{g = 0\}$; Section~\ref{sec:funnel_computation_approach} uses both. Lemma~\ref{lem:s_procedure} is sufficient only. 
In practice, the multiplier degrees are fixed \textit{a priori}, so infeasibility of the resulting SOS program does not imply that the containment or implication fails.

\subsection{Funnel invariance condition}
\label{subsec:funnel_invariance_preliminaries}
Funnels can be certified through a boundary condition on a Lyapunov function $V$. If $V$ increases no more quickly than the level $\rho$ wherever a trajectory touches the funnel boundary, no trajectory can exit. This is the time-varying form of Nagumo's condition for set invariance \citep{nagumo1942lage, blanchini1999set}.

\begin{lemma}[\textsc{Finite-time invariance}]
\label{lem:finite_time_invariance}
Let the deviation dynamics $\bar{\mathbf{f}}$ be locally Lipschitz on $\mathbb{R}^n \times [t_0, t_f]$, and let $V : \mathbb{R}^n \times [t_0, t_f] \to \mathbb{R}$ and $\rho : [t_0, t_f] \to \mathbb{R}$ be continuously differentiable, with sublevel sets $\levelSet(t) \triangleq \{\bar{\stateVector} : V(\bar{\stateVector}, t) \le \rho(t)\}$ compact and $\nabla_{\bar{\stateVector}} V(\bar{\stateVector}, t) \neq \mathbf{0}$ whenever $V(\bar{\stateVector}, t) = \rho(t)$. If, for all $t \in [t_0, t_f]$,
\begingroup
\fontsize{8.5}{10.5}\selectfont
\begin{equation}
    V(\bar{\stateVector}, t) = \rho(t) \implies \dot{V}(\bar{\stateVector}, t) \triangleq \frac{\partial V}{\partial t}(\bar{\stateVector}, t) + \nabla_{\bar{\stateVector}}^\top V(\bar{\stateVector}, t)\, \bar{\mathbf{f}}(\bar{\stateVector}, t) \le \dot{\rho}(t),
    \label{eq:Lyapunov_invariance_condition_funnel_chapter}
\end{equation}
\endgroup
then $\funnel \triangleq \{(t, \stateVector_{nom}(t) + \bar{\stateVector}) : t \in [t_0, t_f],\, \bar{\stateVector} \in \levelSet(t)\}$ is a funnel in the sense of Definition~\ref{def:funnel_introduction}. 
Its cross-sections are $\funnel(t) = \stateVector_{nom}(t) + \levelSet(t) \subseteq \mathbb{R}^n$.
\end{lemma}

\begin{proof}
Augment the state with time, $\mathbf{w} \triangleq (\bar{\stateVector}, t)$, so that $\dot{\mathbf{w}} = (\bar{\mathbf{f}}(\bar{\stateVector}, t), 1)$ has unique solutions, and let $h(\mathbf{w}) \triangleq V(\bar{\stateVector}, t) - \rho(t)$. 
The (Bouligand) tangent cone to $\{h \le 0\}$ at a point is the set of directions that point into the set or along its boundary \citep{blanchini1999set}. Since $\nabla_{\mathbf{w}} h \neq \mathbf{0}$ on $\{h = 0\}$, the tangent cone at a boundary point is the half-space $\{\mathbf{d} : \nabla_{\mathbf{w}} h^\top \mathbf{d} \le 0\}$ of directions that do not increase $h$. Along the dynamics,
$\nabla_{\mathbf{w}} h^\top \dot{\mathbf{w}} = \nabla_{\bar{\stateVector}}^\top V\, \bar{\mathbf{f}} + (\frac{\partial V}{\partial t} - \dot{\rho}) \cdot 1 = \dot{V} - \dot{\rho}$, and \eqref{eq:Lyapunov_invariance_condition_funnel_chapter} implies that $\dot{\mathbf{w}}$ lies in this tangent cone at every boundary point. Nagumo's theorem \citep{nagumo1942lage} then gives forward invariance of $\{h \le 0\}$, i.e., of $\levelSet(t)$ on $[t_0, t_f]$, which is the claim.
\end{proof}

\begin{remark}[\textsc{Applicability}]
For quadratic $V = \bar{\stateVector}^\top \mathbf{P}(t) \bar{\stateVector}$ with $\mathbf{P}(t) \in \mathbb{S}^n_{++}$ continuously differentiable and $\rho(t) > 0$, each $\levelSet(t)$ is an ellipsoid, hence compact, and $\nabla_{\bar{\stateVector}} V = 2\mathbf{P}\bar{\stateVector}$ vanishes only at $\bar{\stateVector} = \mathbf{0}$, which lies inside $\levelSet(t)$. If $\bar{\mathbf{f}}$ is only piecewise continuous in $t$, as under zero-order-hold inputs, Lemma~\ref{lem:finite_time_invariance} applies on each interval, and invariance chains across switching times since trajectories, $V$, and $\rho$ are continuous.
\end{remark}

\subsection{Problem statement}
\label{subsec:funnel_problem}

\begin{problem}[\textsc{Minimum-volume certified funnel}]
\label{prob:min_volume_funnel}
Given dynamics \eqref{eq:nonlinear_system_dynamics_funnel_chapter}, boundary states $\stateVector_i$, $\stateVector_f$, and inlet and outlet shape matrices $\mathbf{M}_I, \mathbf{M}_T \in \mathbb{S}^n_{++}$ with $\mathbf{M}_I \preceq \mathbf{M}_T$, compute a nominal trajectory from $\stateVector_i$ to $\stateVector_f$, a tracking feedback control law, and a funnel $\funnel$ certified by Lemma~\ref{lem:finite_time_invariance}, such that
\begin{equation}
    \ellipsoidalSet_I \subseteq \levelSet(t_0), \qquad \levelSet(t_f) \subseteq \ellipsoidalSet_T,
    \label{eq:inlet_outlet_containment}
\end{equation}
while minimizing the funnel volume $\int_{t_0}^{t_f} \mathrm{vol}\big(\levelSet(t)\big)\, dt$.
\end{problem}

An optimal solution to Problem~\ref{prob:min_volume_funnel} thus yields the tightest (in volume) outer approximation of the reachable tube of $\ellipsoidalSet_I$ among those certified by Lemma~\ref{lem:finite_time_invariance}.

\begin{remark}[\textsc{Composability by construction}]
\label{rem:composability_by_construction}
Since $\mathbf{M}_I \preceq \mathbf{M}_T$ is equivalent to $\ellipsoidalSet_T \subseteq \ellipsoidalSet_I$, two funnels $\funnel^a$, $\funnel^b$ that solve Problem~\ref{prob:min_volume_funnel} with common $(\mathbf{M}_I, \mathbf{M}_T)$ and whose nominal trajectories meet, $\stateVector^a_{nom}(t^a_f) = \stateVector^b_{nom}(t^b_0)$, satisfy $\levelSet^a(t^a_f) \subseteq \ellipsoidalSet_T \subseteq \ellipsoidalSet_I \subseteq \levelSet^b(t^b_0)$; translating by the common endpoint gives $\funnel^a(t^a_f) \subseteq \funnel^b(t^b_0)$, so the pair is \textit{sequentially composable} \eqref{eq:composability_stricter}. A library of such funnels is therefore composable wherever nominal endpoints coincide, with no pairwise checks needed.
\end{remark}

Problem~\ref{prob:min_volume_funnel} is non-convex and infinite-dimensional. Our method (Section~\ref{sec:funnel_computation_approach}) makes it tractable by fixing the nominal trajectory and feedback law before certification, and by restricting $V$ to polynomials in $\bar{\stateVector}$ with $(V, \rho)$ sampled at discrete times (knot points), which turns certification into a single bilinear SOS program, solved as a sequence of SDPs. 

\section{Method}
\label{sec:funnel_computation_approach}
\begin{figure}[t]
\centering
\begin{tikzpicture}[
    font=\footnotesize,
    blk/.style={draw, rounded corners=2pt, align=center,
                minimum width=3.3cm, minimum height=1.0cm, inner sep=3pt},
    arr/.style={-{Stealth[length=2mm]}, thick},
    node distance=1.65cm and 1.75cm]
  \node[blk] (traj) {Nominal trajectory\\(direct collocation)};
  \node[blk, right=of traj] (lqr) {Feedback synthesis\\(TVLQR)};
  \node[blk, below=of lqr] (poly) {Polynomialization\\(Taylor expansion)};
  \node[blk, below=of traj] (sos) {SOS certification\\(bilinear alternation)};
  \draw[arr] ([yshift=0.7cm]traj.north) node[above] {$\stateVector_i,\, \stateVector_f,\, \mathbf{f}$} -- (traj.north);
    \draw[arr] ([yshift=0.7cm]sos.north) node[above] {$\mathbf{M}_I,\, \mathbf{M}_T$} -- (sos.north);
  \draw[arr] (traj) -- node[above] {$\stateVector_{nom}, \inputVector_{nom}$} (lqr);
  \draw[arr] (lqr) -- node[right] {$\mathbf{K}_k$} (poly);
  \draw[arr] (poly) -- node[above] {$\widetilde{\mathbf{f}}_k$} (sos);
  \draw[arr, dashed] (lqr.south west) -- node[pos=0.45, above, sloped] {$\mathbf{S}_k$ (warm start)} (sos.north east);
  \draw[arr] (sos.south) -- ([yshift=-0.7cm]sos.south) node[below] {$\{(V_k, \rho_k)\}_{k=0}^{K}$};
\end{tikzpicture}
\caption{Funnel computation pipeline. The TVLQR cost-to-go matrices $\mathbf{S}_k$ warm-start the bilinear alternation of SOS program solve.}
\label{fig:funnel_pipeline}
\end{figure}
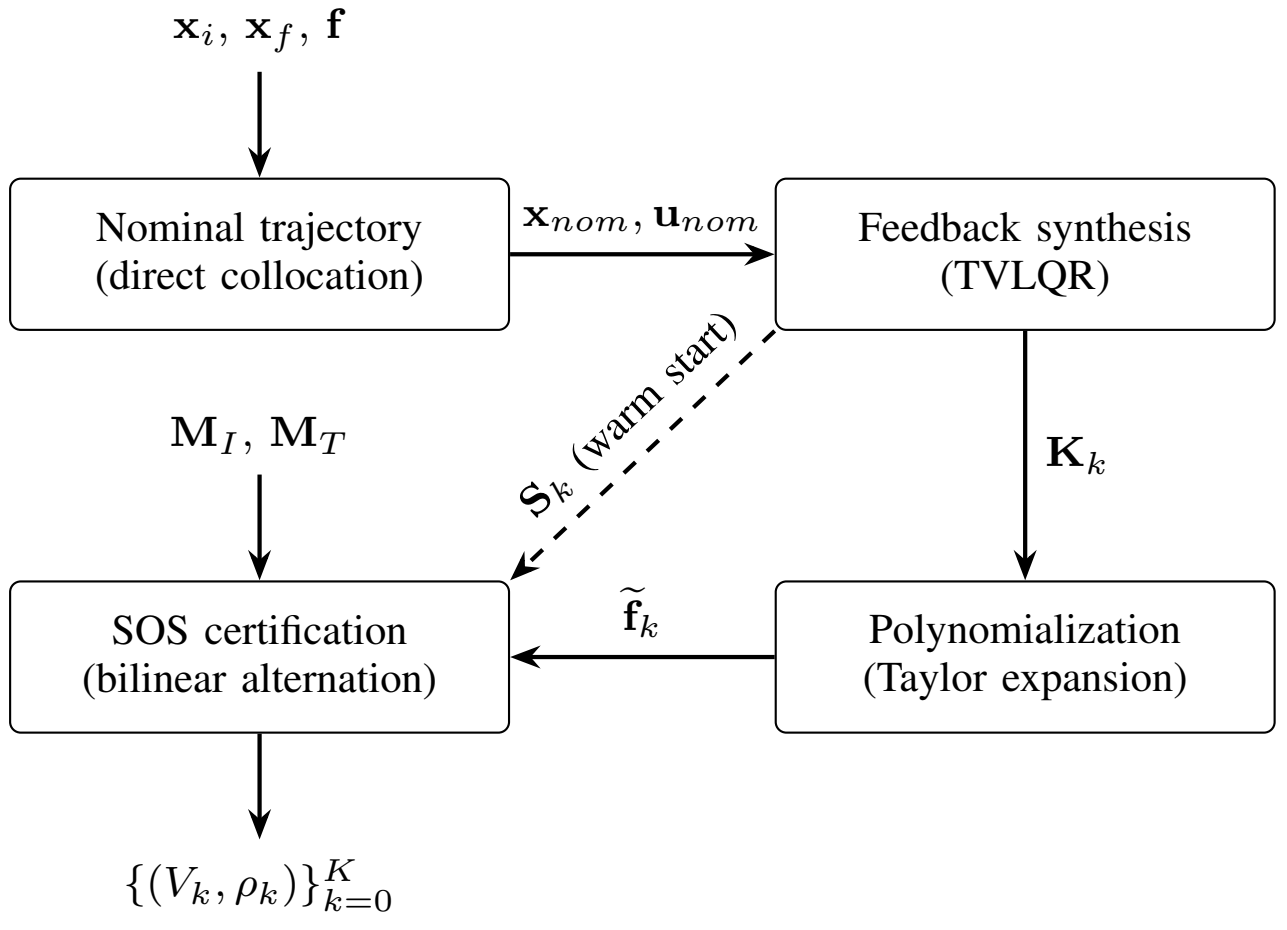

Following an approach similar to \citep{jaffar2022pip}, we solve Problem~\ref{prob:min_volume_funnel} in four sequential steps (see \figref{fig:funnel_pipeline}), parameterizing the funnel by sublevel sets of polynomial Lyapunov functions at knot points. The steps for computing these funnels are outlined below.

\subsection{Optimization-based nominal trajectory generation}
\label{subsec:trajectory_optimization_funnel_chapter}

We solve a nonlinear program (NLP) to compute nominal trajectory $\stateVector_{nom}$ and feedforward input tape $\inputVector_{nom}$ that drive the system from a specified initial state $\stateVector_i$ to final state $\stateVector_f$. We implement a direct collocation-based approach with trapezoidal integration constraints to solve the optimization problem \citep{kelly2017introduction}.
Let $t_k \triangleq k\,\Delta T$ denote the time at knot point $k \in \mathcal{K} \triangleq \{0, \dots, K\}$ where $K$ is user-specified, so that $t_0 = 0$ and $t_f \equiv t_K$. Let the input be held constant over each interval, $\inputVector(t) = \inputVector_k$ for
$t \in [t_k, t_{k+1})$. The nominal trajectory is obtained by solving
\begingroup
\fontsize{9}{10.5}\selectfont
\begin{subequations}
\label{eq:nominal_trajectory_optimization}
\begin{align}
  \underset{\stateVector_{0:K},\,\inputVector_{0:K-1},\,\Delta T}{\text{minimize}}
    \quad & \sum_{k=0}^{K-1} \inputVector_k^\top \mathbf{R}_{\mathrm{nom}}\,
            \inputVector_k + w_t\, t_K
      \label{eq:traj_opt_cost} \\
  \text{subject to} \:\:\: \quad \quad
    & \stateVector_{k+1} \!= \stateVector_k
      \!+\! \tfrac{\Delta T}{2}\big(\boldsymbol{\kappa}_1 \!+\! \boldsymbol{\kappa}_2\big),
      \, k \in \mathcal{K} \!\setminus\! \{K\},
      \label{eq:traj_opt_trapezoidal} \\
    & \stateVector_0 = \stateVector_i, \quad \stateVector_K = \stateVector_f,
      \label{eq:traj_opt_bc} \\
    & \inputVector_k \in \mathcal{U}_{\mathrm{feas}}, \;\, k \in \mathcal{K} \setminus \{K\},
      \label{eq:traj_opt_feas_input} \\
    & 0 < \Delta T \le \Delta T_{\max},
      \label{eq:traj_opt_dt}
\end{align}
\end{subequations}
\endgroup
where $\boldsymbol{\kappa}_1 = \mathbf{f}(\stateVector_k, \inputVector_k)$ and $\boldsymbol{\kappa}_2 = \mathbf{f}\big(\stateVector_{k+1},\, \inputVector_k\big)$ in \eqref{eq:traj_opt_trapezoidal}, and $\mathcal{U}_{\mathrm{feas}}$ characterizes input limits. $\mathbf{R}_{\mathrm{nom}} \succ 0$ penalizes control effort, weight coefficient $w_t > 0$ trades off
control effort against trajectory duration $t_K = K \Delta T$. Since $K$ is fixed and $\Delta T$ is a decision variable, \eqref{eq:nominal_trajectory_optimization} is a free final-time problem.
The solution to \eqref{eq:nominal_trajectory_optimization} yields the nominal trajectory $\stateVector_{nom} = \{\stateVector_k\}_{k=0}^{K}$ and the feedforward input tape $\inputVector_{nom} = \{\inputVector_k\}_{k=0}^{K-1}$. The trapezoidal constraints enforce the dynamics only approximately; we neglect this collocation defect and treat $(\stateVector_{nom}, \inputVector_{nom})$ as an exact solution of \eqref{eq:nonlinear_system_dynamics_funnel_chapter}.

\subsection{Feedback controller design via TVLQR}
\label{subsec:tvlqr_synthesis_funnel_chapter}

To reject deviations from the nominal trajectory, we augment the feedforward input with a time-varying linear quadratic regulator (TVLQR). We linearize \eqref{eq:nonlinear_system_dynamics_funnel_chapter} at each knot point $k$, $\mathbf{A}_k \triangleq \partial \mathbf{f}/\partial \stateVector$ and $\mathbf{B}_k \triangleq \partial \mathbf{f}/\partial \inputVector$ evaluated at $({\stateVector_{nom}}_k, {\inputVector_{nom}}_k)$, and interpolate linearly in time to obtain $\mathbf{A}(t)$ and $\mathbf{B}(t)$ on $[t_0, t_f]$. With state and input weights $\mathbf{Q} \succeq 0$, $\mathbf{R} \succ 0$ and terminal weight $\mathbf{Q}_f \succeq 0$, the cost-to-go matrix $\mathbf{S}(t)$ solves the Riccati differential equation
\begin{equation}
    -\dot{\mathbf{S}} = \mathbf{A}^\top\mathbf{S} + \mathbf{S}\mathbf{A} - \mathbf{S}\mathbf{B}\mathbf{R}^{-1}\mathbf{B}^\top\mathbf{S} + \mathbf{Q} , \qquad \mathbf{S}(t_f) = \mathbf{Q}_f,
    \label{eq:riccati_ode}
\end{equation}
which we integrate backward with a stiff solver (MATLAB \texttt{ode15s}); the continuous-time formulation avoids sensitivity of the Riccati equation solution to the sampling time. The feedback gain is $\mathbf{K}(t) = \mathbf{R}^{-1}\mathbf{B}^\top(t)\mathbf{S}(t)$, and we extract the matrix values at the knot points: $\mathbf{S}_k \triangleq \mathbf{S}(t_k)$ and $\mathbf{K}_k \triangleq \mathbf{K}(t_k)$. The closed-loop control law is
\begingroup
\begin{equation}
    \inputVector_k = {\inputVector_{nom}}_k - \mathbf{K}_k(\stateVector - {\stateVector_{nom}}_k), \quad \forall k \in \{0, \dots, K-1\},
    \label{eq:tvlqr_control_law}
\end{equation}
\endgroup
with $\inputVector_K \triangleq \inputVector_{K-1}$ so that the control law is defined at every knot. Note that \eqref{eq:traj_opt_feas_input} bounds $\inputVector_{nom}$, but the feedback term can exceed $\mathcal{U}_{\mathrm{feas}}$ inside the funnel certificate (rare in practice).

\subsection{Deviation dynamics and polynomialization}
\label{subsec:deviation_dynamics_funnel_chapter}

Substituting \eqref{eq:tvlqr_control_law} into \eqref{eq:nonlinear_system_dynamics_funnel_chapter} and subtracting the nominal dynamics yields the closed-loop deviation dynamics. At each knot point $k \in \mathcal{K}$,
\begingroup
\fontsize{9.5}{11}\selectfont
\begin{align}
\begin{split}
\label{eq:deviation_dynamics}
    \dot{\bar{\stateVector}} &= \bar{\mathbf{f}}_k(\bar{\stateVector}) \triangleq \bar{\mathbf{f}}(\bar{\stateVector}, t_k) \\
    &= \mathbf{f}\big({\stateVector_{nom}}_k + \bar{\stateVector},\, {\inputVector_{nom}}_k - \mathbf{K}_k\bar{\stateVector}\big) - \mathbf{f}({\stateVector_{nom}}_k, {\inputVector_{nom}}_k).
\end{split}
\end{align}
Recall state deviation, $\bar{\stateVector} = \stateVector - \stateVector_{nom}$. Since SOS programming requires polynomial functions, we replace $\bar{\mathbf{f}}_k$ with its third-order Taylor polynomial $\widetilde{\mathbf{f}}_k$ about $\bar{\stateVector} = \mathbf{0}$, following \citet{Tobenkin.etal.IFAC11, Majumdar.Tedrake.IJRR17}:
\begin{equation}
    \dot{\bar{\stateVector}} \approx \widetilde{\mathbf{f}}_k(\bar{\stateVector}), 
    \quad k \in \mathcal{K} = \{0,\dots,K\}.\label{eq:deviation_dynamics_approximated}
\end{equation}
\endgroup
Note that invariance is checked on the funnel boundary, away from the origin, where polynomial approximation errors are large and grow with funnel size. The invariance certificates obtained from SOS programming (Section~\ref{subsec:invariance_sos_certification_funnel_chapter}) holds for $\widetilde{\mathbf{f}}_k$ only. Section~\ref{sec:results_funnel_chapter} empirically verifies the certificates against $\bar{\mathbf{f}}_k$ on states sampled from funnel boundary.

\subsection{Invariance certification via SOS programming}
\label{subsec:invariance_sos_certification_funnel_chapter}

We parametrize the funnel at the knot points by $\levelSet_k \triangleq \{\bar{\stateVector} : V_k(\bar{\stateVector}) \le \rho_k\}$, $k \in \mathcal{K}$, with polynomial $V_k$ and scalar $\rho_k > 0$, so that $\funnel(t_k) = \stateVector_k + \levelSet_k$. Replacing the time derivatives in \eqref{eq:Lyapunov_invariance_condition_funnel_chapter} by backward differences and $\bar{\mathbf{f}}$ by $\widetilde{\mathbf{f}}_k$, and omitting the argument $\bar{\stateVector}$ in $V_k$ for brevity, yields the sampled invariance condition:
\begingroup
\fontsize{9}{10.5}\selectfont
\begin{align}
\begin{split}
    &V_k = \rho_k \implies \dot{V}_k \le \dot{\rho}_k, \qquad k \in \mathcal{K} \setminus \{0\}, \\
    &\dot{V}_k \approx \frac{V_k - V_{k-1}}{t_k - t_{k-1}} + \nabla_{\bar{\stateVector}}^\top V_k\, \widetilde{\mathbf{f}}_k, \qquad \dot{\rho}_k \approx \frac{\rho_k - \rho_{k-1}}{t_k - t_{k-1}}.
\end{split}
\label{eq:Lyapunov_invariance_condition_discretized}
\end{align}
\endgroup
By Lemma~\ref{lem:s_procedure}(b) with $p = \dot{\rho}_k - \dot{V}_k$ and $g = \rho_k - V_k$, condition \eqref{eq:Lyapunov_invariance_condition_discretized} holds if there exist polynomials $\lambda_k \in \mathbb{R}[\bar{\stateVector}]$ such that
\begin{equation}
    \dot{\rho}_k - \dot{V}_k - \lambda_k\, (\rho_k - V_k) \in \Sigma[\bar{\stateVector}].
    \label{eq:S_procedure_invariance_constraint}
\end{equation}
The multiplier $\lambda_k$ is sign-free because the implication is required only on the level set $\{V_k = \rho_k\}$. As remarked earlier in Section~\ref{subsec:funnel_s_procedure}, condition \eqref{eq:S_procedure_invariance_constraint} is sufficient but not necessary, bringing in conservativeness to our proposed method.

With $\Sigma \triangleq \Sigma[\bar{\stateVector}]$ (cone of SOS polynomials), the minimum-volume certified funnel (Problem~\ref{prob:min_volume_funnel}) is computed by the following SOS program.
\begingroup
\fontsize{9}{10.5}\selectfont
\begin{subequations}
\label{eq:SOS_optimization_problem}
\begin{align}
  \min_{\substack{V_{0:K},\;\rho_{0:K},\\ \lambda_{1:K},\;\mu_I,\;\mu_T}} \;\;
    & \sum_{k=0}^{K} \rho_k \label{eq:sos_cost} \\
  \text{s.t.} \qquad \,
    & V_k - \epsilon\,\bar{\stateVector}^\top \bar{\stateVector} \in \Sigma, \;\;
      V_k(\mathbf{0}) = 0, \;\; k \in \mathcal{K}, \label{eq:sos_pd} \\
    & \rho_k > 0, \;\; k \in \mathcal{K} \label{eq:sos_rho_pos} \\
    & \text{invariance condition \eqref{eq:S_procedure_invariance_constraint}}, \;\;
      k \in \mathcal{K} \setminus \{0\}, \label{eq:sos_invariance} \\
    & (\rho_0 - V_0) - \mu_I\, \big(1 - \bar{\stateVector}^\top \mathbf{M}_I \bar{\stateVector}\big) \in \Sigma, \; \mu_I \in \Sigma,
      \label{eq:sos_initial} \\
    & \big(1 - \bar{\stateVector}^\top \mathbf{M}_T \bar{\stateVector}\big) - \mu_T\,(\rho_K \!-\! V_K) \!\in \Sigma, \; \mu_T \!\in \Sigma,
      \label{eq:sos_terminal} \\
    & V_k(\mathbf{1}_n) = V_{\mathrm{guess}}(\mathbf{1}_n), \;\; k \in \mathcal{K}\,. \label{eq:sos_normalization}
\end{align}
\end{subequations}
\endgroup

Constraint \eqref{eq:sos_pd} enforces positive definiteness of each $V_k$ for a small $\epsilon > 0$, and hence compactness of $\levelSet_k$. By Lemma~\ref{lem:s_procedure}(a), \eqref{eq:sos_initial} certifies $\ellipsoidalSet_I \subseteq \levelSet_0$ and \eqref{eq:sos_terminal} certifies $\levelSet_K \subseteq \ellipsoidalSet_T$; their multipliers must be SOS because they certify containment of sets. 
Together with \eqref{eq:sos_invariance}, the program enforces the inlet and outlet conditions \eqref{eq:inlet_outlet_containment} and a sampled-time approximation of invariance (Lemma~\ref{lem:finite_time_invariance}) for the approximate dynamics $\widetilde{\mathbf{f}}_k$.

In our implementation $V_k$ is quadratic and $\mu_I, \mu_T$ are nonnegative scalars, so the polynomials in \eqref{eq:sos_initial}--\eqref{eq:sos_terminal} are quadratic, for which nonnegativity and SOS coincide (Section~\ref{subsec:funnel_sos}). Since the origin lies in the interior of $\ellipsoidalSet_I$ and of $\levelSet_K$, the S-lemma \citep{boyd1994linear} makes both certificates necessary as well as sufficient. The conservatism of the SOS relaxation therefore resides entirely in the invariance constraint \eqref{eq:sos_invariance}.

Condition \eqref{eq:Lyapunov_invariance_condition_funnel_chapter} is invariant to time-varying rescaling $(V, \rho) \mapsto (\alpha(t)V, \alpha(t)\rho)$, $\alpha(t) > 0$, which leaves every $\levelSet(t)$ unchanged but alters the objective \eqref{eq:sos_cost}. The normalization \eqref{eq:sos_normalization} fixes this scale at each knot. Even so, $\rho_k$ does not determine $\mathrm{vol}(\levelSet_k)$, which also depends on the shape of $V_k$; \eqref{eq:sos_cost} is therefore a heuristic proxy for funnel volume, and we report the exact volume in all results (Section~\ref{sec:results_funnel_chapter}).

SOS program~\eqref{eq:SOS_optimization_problem} is bilinear (non-convex) in the decision variables through the products $\lambda_k V_k$, $\lambda_k \rho_k$, $\mu_T V_K$, and $\mu_T \rho_K$. We solve it by bilinear alternation (Algorithm~\ref{alg:bilinear_alternation}), which alternates between two SDPs: 

\begin{enumerate}
    \item L-Step \textit{(Feasibility):} Fix $(V_k, \rho_k)$ and search for certifying multipliers $\lambda_k$, $\mu_I$ and $\mu_T$, subject to \eqref{eq:sos_invariance}--\eqref{eq:sos_terminal}.
    \item V-Step \textit{(Optimization):} Fix multipliers obtained from \mbox{L-step} to minimize \eqref{eq:sos_cost} over \mbox{($V_k$, $\rho_k$)}, subject to all constraints \eqref{eq:sos_pd}--\eqref{eq:sos_normalization}.
\end{enumerate}

Algorithm~\ref{alg:bilinear_alternation} terminates when the relative decrease in the objective falls below a tolerance $\delta$, or after $\mathrm{iter}_{\max}$ iterations. 
The SDP at each V-step is feasible, since the preceding L-step certifies it, so the objective is non-increasing and, being bounded below, converges. For the same reason, L-step can fail only at the first iteration, which makes feasibility of the initial guess the critical input. Convergence is of the objective value only, with no guarantee of global or local optimality.

\paragraph{Funnel volume}
With quadratic $V_k(\bar{\stateVector}) = \bar{\stateVector}^\top \mathbf{P}_k \bar{\stateVector}$, $\mathbf{P}_k \in \mathbb{S}^n_{++}$, each $\levelSet_k = \{\bar{\stateVector} : \bar{\stateVector}^\top (\mathbf{P}_k/\rho_k)\, \bar{\stateVector} \le 1\}$ is an ellipsoid of volume \citep{boyd1994linear}
\begin{equation}
    \mathrm{vol}(\levelSet_k) = \omega_n \sqrt{\rho_k^n / \det \mathbf{P}_k},
    \label{eq:ellipsoid_volume}
\end{equation}
where $\omega_n$ is the volume of the unit ball in $\mathbb{R}^n$. We measure funnel volume by
\begin{equation}
    \mathrm{vol}(\funnel) \triangleq \Delta T \sum_{k \in \mathcal{K}} \mathrm{vol}(\levelSet_k),
    \label{eq:approximate_funnel_volume}
\end{equation}
which approximates the objective $\int_{t_0}^{t_f} \mathrm{vol}(\levelSet(t))\, dt$ of Problem~\ref{prob:min_volume_funnel}. 

\begin{remark}[\textsc{Initial Guess for $V$ and $\rho$}]
    To initialize the bilinear alternation procedure, we must specify an \textit{initial guess} for the Lyapunov candidate function $V_k(\bar{\stateVector})$ and the level set value $\rho_k$ at all $k \in \{0, \dots, K\}$. Similar to \citep{Tobenkin.etal.IFAC11}, we use the following guesses:
    \begin{align}
    \begin{split}
        &{V_{\mathrm{guess}}}_k(\bar{\stateVector}) = \bar{\stateVector}^\top \mathbf{S}_k \bar{\stateVector}\:, \\
        &{\rho_{\mathrm{guess}}}_k = \rho_{\mathrm{g}} e^{c(t_k - t_K)/(t_0 - t_K)}.
    \end{split}
    \label{eq:V_and_rho_initial_guess}
    \end{align}
    At $k = K$ the exponent vanishes, so $\rho_{\mathrm{g}}$ sets the outlet level, and the ratio ${\rho_{\mathrm{guess}}}_0 / {\rho_{\mathrm{guess}}}_K = e^{c}$. $c > 0$ yields a guess whose level decreases with time. 
    
    $\mathbf{S}_k$ is the cost-to-go matrix obtained by solving the Riccati differential equation \eqref{eq:riccati_ode} in TVLQR synthesis, while $\rho_{\mathrm{g}}$ and $c$ are tunable hyperparameters. To ensure the guesses are good for initial feasibility, we prioritize obtaining numerically well-conditioned $\mathbf{S}_k$. We use Bryson's rule for coarse-tuning $\mathbf{Q}$ and $\mathbf{R}$ matrices \citep{bryson2018applied}, followed by trial-and-error fine-tuning based on condition number of $\mathbf{S}_k$ (we target $\mathrm{cond}(\mathbf{S}_k) < 10^3$, an empirical ideal threshold). 
\end{remark}

\begin{algorithm}[t]
 \caption{\subroutine{bilinearAlternation}\,($\widetilde{\mathbf{f}}_{1:K}, \mathbf{M}_I, \mathbf{M}_T$)}
 \begin{algorithmic}[1]
    \State $(V_k, \rho_k) \gets ({V_{\mathrm{guess}}}_k, {\rho_{\mathrm{guess}}}_k)$, $k \in \mathcal{K}$ \Comment{initial guess \eqref{eq:V_and_rho_initial_guess}}
    \For{$\mathrm{iter} = 1, \dots, \mathrm{iter}_{\max}$}
        \State find $\lambda_{1:K}, \mu_I, \mu_T$ s.t.\ \eqref{eq:sos_invariance}--\eqref{eq:sos_terminal} \Comment{L-step} \label{algline:L_step}
        \If{L-step infeasible} 
            \State \Return failure
        \EndIf
        \State solve \eqref{eq:SOS_optimization_problem} for $(V_k, \rho_k)$ \Comment{V-step} \label{algline:V_step}
        \State $J_\mathrm{iter} \gets \sum_{k \in \mathcal{K}} \rho_k$
        \If{$\mathrm{iter} > 1$ \textbf{and} $(J_{\mathrm{iter}-1} - J_\mathrm{iter})/J_\mathrm{iter} < \delta$}
           \State \textbf{break} \Comment{converged}
        \EndIf
    \EndFor
    \State \Return $(V_k, \rho_k)_{k \in \mathcal{K}}$
  \end{algorithmic}
  \label{alg:bilinear_alternation}
\end{algorithm}

\begin{table*}[t!]
\caption{Unicycle problem and Solver parameters}
\label{tab:setup_funnel_chapter}
\centering
\begin{tabular}{ll}
\hline
Parameter & Value \\
\hline 
\rule{0pt}{2.5ex}Dynamics & $\dot{x} = v\cos\theta,\; \dot{y} = v\sin\theta,\; \dot{\theta} = \omega$ \\
$\stateVector_i$, $\stateVector_f$ & $[0\ 0\ \pi/2]^\top$, $[2\ 4\ \pi/2]^\top$ \\
$\mathcal{U}_{\mathrm{feas}}$ & $|v| \le$ 0.75\,m/s, $|\omega| \le$ $\pi$/2\,rad/s \\
$\Delta T_{\max}$, $K\!+\!1$ & 0.25\,s, 40  \\
$\mathbf{R}_{\mathrm{nom}}$, $w_t$ & diag([0.1, 0.1]), 1 \\  
$\mathbf{Q}$, $\mathbf{R}$ & diag([1, 1, 0.1]), diag([1, 0.5]) \\
$\mathbf{M}_I$, $\mathbf{M}_T$ & see \eqref{eq:inlet_outlet_setup} \\
$\deg V_k$, $\deg \widetilde{\mathbf{f}}_k$, $\deg \lambda_k$, $\deg \mu_{I,T}$ & 2, 3, 6, 0 \\
$\mathrm{iter}_{\max}$, $\delta$, $\epsilon$ & 20, 0.05, 10$^{-6}$\\
Solver & SOSTOOLS v4.00 + MOSEK v10.2.8, MATLAB R2023a \\ 
Hardware and OS & Intel Core i7-10750H (6 cores, 2.6\,GHz), 32\,GB RAM, Ubuntu 20.04 \\
\hline
\end{tabular}
\end{table*}

\begin{remark}[\textsc{Convergent funnels}]
    To make the computed funnels composable, we choose $\mathbf{M}_I \preceq \mathbf{M}_T$, i.e., $\ellipsoidalSet_T \subseteq \ellipsoidalSet_I$ (see Remark~\ref{rem:composability_by_construction}), and warm-start Algorithm~\ref{alg:bilinear_alternation} with a convergent guess ($c > 0$), selecting $(\rho_{\mathrm{g}}, c)$ in \eqref{eq:V_and_rho_initial_guess} by a sweep over initial feasibility. A decreasing $\rho_{\mathrm{guess}}$ does not by itself imply decreasing volume, since $\mathbf{S}_k$ also varies along the trajectory; the containment constraints \eqref{eq:sos_initial}--\eqref{eq:sos_terminal} are what enforce shrinking volume behavior.
\end{remark}

\section{Numerical results}
\label{sec:results_funnel_chapter}

All results in this section use a unicycle system with state $\stateVector = (x, y, \theta)$, input $\inputVector = (v, \omega)$, and dynamics $\dot{x} = v\cos\theta$, $\dot{y} = v\sin\theta$, $\dot{\theta} = \omega$. Additional systems, cart-pole ($n = 4$) and quadrotor ($n = 12$) are considered only for computation timing study  (Section~\ref{subsec:results_discussion_funnel_chapter}). The implementation scripts are available as part of our \texttt{computeSOSFunnels}\footnote{\href{https://github.com/khalid2696/computeSOSFunnels/tree/unicycle}{\texttt{github.com/khalid2696/computeSOSFunnels/tree/unicycle}}} package.

\subsection{Setup}
\label{subsec:results_setup_funnel_chapter}

Table~\ref{tab:setup_funnel_chapter} lists all problem and solver parameters. The nominal trajectory from $\stateVector_i$ to $\stateVector_f$ uses $K = 39$ intervals (40 knots). The user-specified inlet and outlet sets are derived from the TVLQR cost-to-go matrices $\mathbf{S}$ and the default initial guess hyperparameters $(\rho_{\mathrm{g}}, c) = (0.05, 1)$,
\begin{equation}
    \mathbf{M}_T = 0.8\, \mathbf{S}_K / \rho_{\mathrm{g}}, \qquad \mathbf{M}_I = 1.25\, \mathbf{S}_0 / (\rho_{\mathrm{g}} e^{c}),
    \label{eq:inlet_outlet_setup}
\end{equation}
and held fixed in all experiments. They satisfy $\lambda_{\min}(\mathbf{M}_T - \mathbf{M}_I) = 0.7 > 0$ (i.e., $\mathbf{M}_T \succ \mathbf{M}_I \implies \ellipsoidalSet_T \subseteq \ellipsoidalSet_I$), so every computed funnel is sequentially composable by construction (Remark~\ref{rem:composability_by_construction}). We implement Algorithm~\ref{alg:bilinear_alternation} in SOSTOOLS \citep{SOSTOOLS} with MOSEK \citep{MOSEK} as the SDP solver. 
All funnel volume metrics follow \eqref{eq:ellipsoid_volume}--\eqref{eq:approximate_funnel_volume}.



\subsection{Representative funnel}
\label{subsec:results_representative_funnel_chapter}

\begin{figure*}[t]
\centering
\begin{minipage}{0.24\textwidth}\centering
  \includegraphics[width=\linewidth]{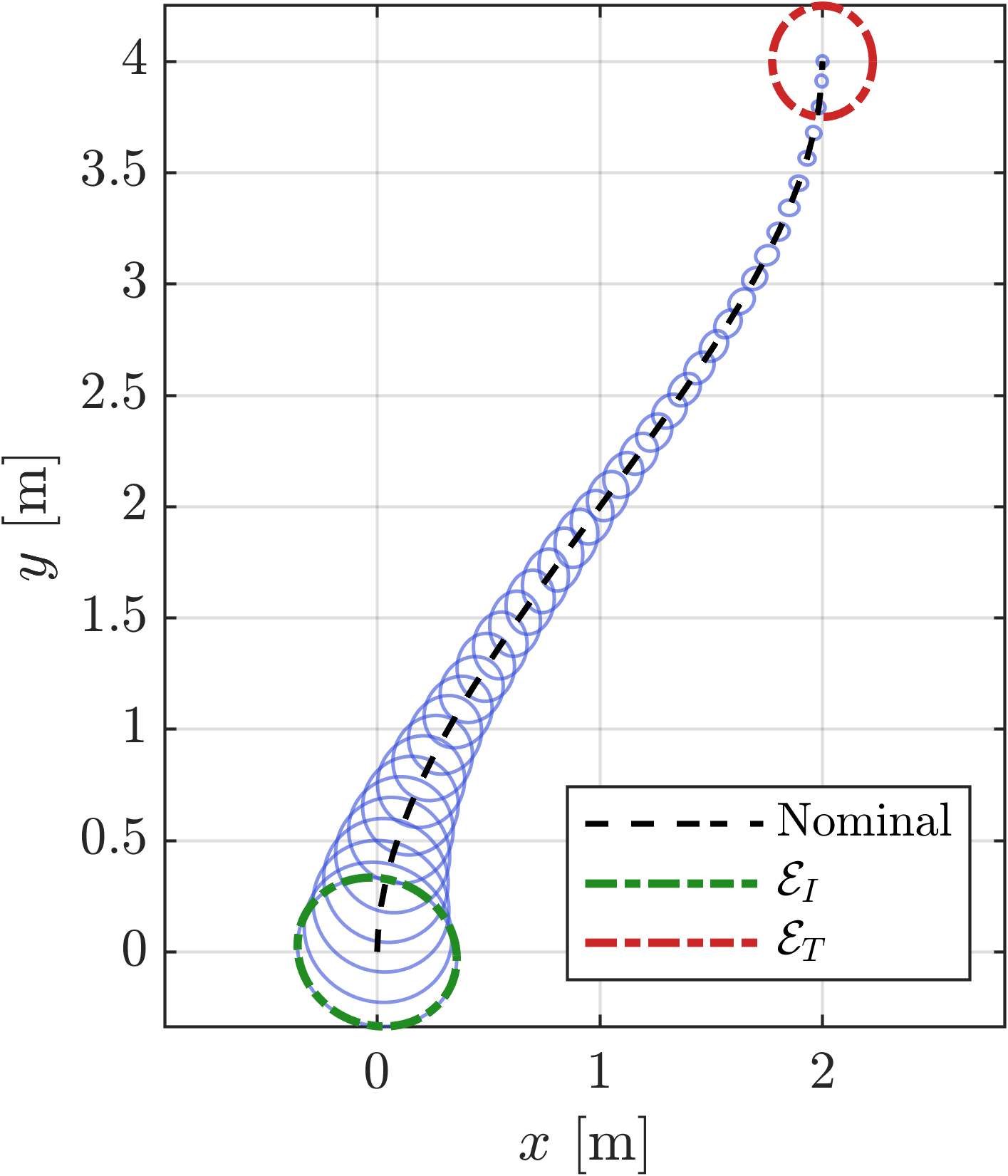}\\(a)
\end{minipage} \hspace{0.2em}
\begin{minipage}{0.29\textwidth}\centering
  \includegraphics[width=\linewidth]{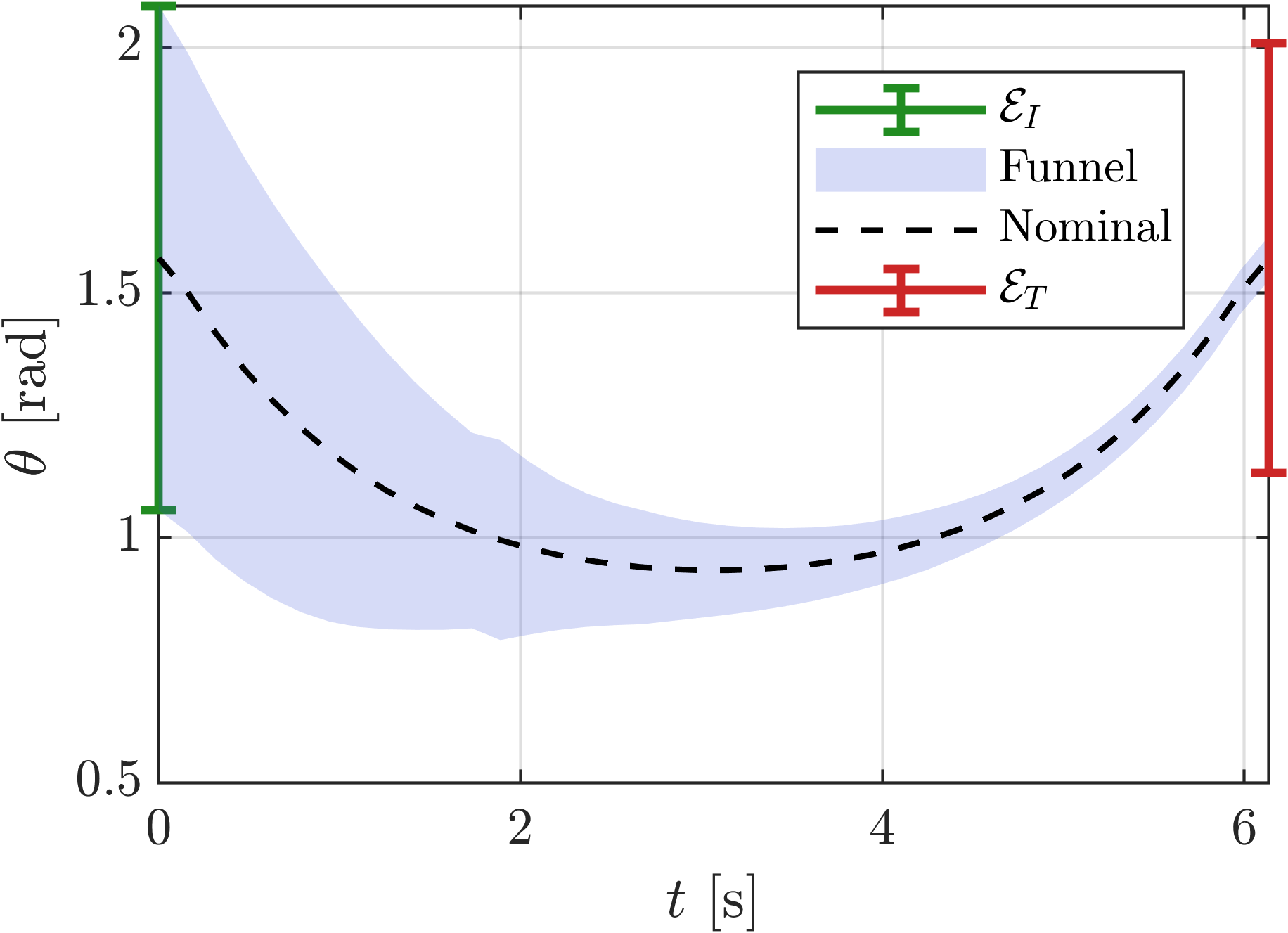}\\(b)
\end{minipage} \hspace{0.2em}
\begin{minipage}{0.31\textwidth}\centering
  \includegraphics[width=\linewidth]{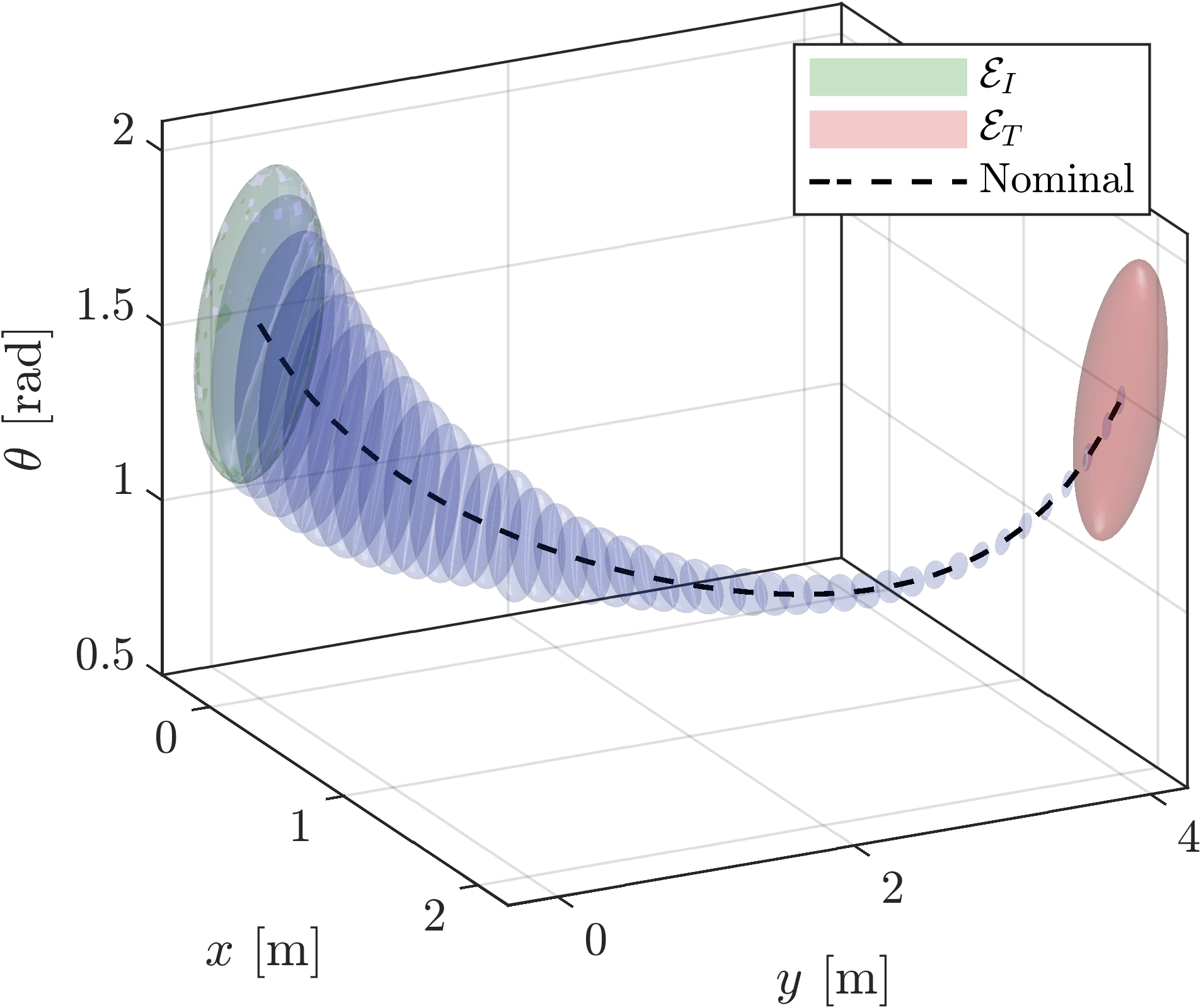}\\(c)
\end{minipage}
\caption{Funnel computed by Algorithm~\ref{alg:bilinear_alternation} for a unicycle system, terminated at $\mathrm{iter} = 4$. (a) Projections of the cross-sections $\funnel(t_k)$ onto $xy$-plane, with the inlet set $\ellipsoidalSet_I$ and the outlet set $\ellipsoidalSet_T$. (b) Projection onto $\theta$-space over time; vertical bars mark the $\theta$-extent of $\ellipsoidalSet_I$ and $\ellipsoidalSet_T$. (c) Funnel cross-sections in $(x, y, \theta)$.}
\label{fig:representative_funnel_chapter}
\end{figure*}

Fig.~\ref{fig:representative_funnel_chapter} illustrates the funnel output by Algorithm~\ref{alg:bilinear_alternation} on the default problem (Table~\ref{tab:setup_funnel_chapter}), which terminates at $\mathrm{iter} = 4$ (convergence criterion met). The inlet constraint is active: the computed inlet $\levelSet_0$ matches $\ellipsoidalSet_I$ in volume to within $10^{-4}$, as expected, since \eqref{eq:sos_cost} penalizes $\rho_0$ and \eqref{eq:sos_initial} bounds it from below. The outlet constraint is far from active: $\levelSet_K$ occupies 0.11\% of the volume of $\ellipsoidalSet_T$. The resulting inlet-to-outlet volume ratio is 2205, nearly three orders of magnitude above the floor of 2.47 that $\mathbf{M}_I \preceq \mathbf{M}_T$ alone enforces. The funnel thus contracts well beyond what composability requires, leaving a wide margin for hand-off to a successor funnel.

\subsection{Hyperparameter sweep}
\label{subsec:results_sweep_funnel_chapter}

Table~\ref{tab:sweep_funnel_chapter} sweeps the initial-guess hyperparameters $(\rho_{\mathrm{g}}, c)$ in \eqref{eq:V_and_rho_initial_guess} with fixed $\ellipsoidalSet_I$ and $\ellipsoidalSet_T$ and a fixed budget of $4$ iterations for Algorithm~\ref{alg:bilinear_alternation} termination. Of the considered 16 initial guesses in this problem, seven violate inlet or outlet containment (denoted by `C'). Since $V_{\mathrm{guess}}$ shares the shapes of $\mathbf{M}_I$ and $\mathbf{M}_T$, these failures are decidable even before solving the SOS program \eqref{eq:SOS_optimization_problem}. 
Of the remaining nine, the four that fail feasibility check in the initial L-step (marked as `--') are those with the largest guessed inlet level, $\rho_{\mathrm{g}} e^{c} \ge 0.36$, whereas all guesses with $\rho_{\mathrm{g}} e^{c} \le 0.26$ succeed --- a funnel guess that is too wide at the inlet cannot be certified locally. 
Among the five feasible guesses, the funnel volume varies by 4.6\% and the inlet-to-outlet ratio by 26\%. So the warm start has little effect on funnel volume but a noticeable effect on how strongly the funnel contracts.

\begin{table}[t]
\caption{Sweep over the initial guess hyperparameters $(\rho_{\mathrm{g}}, c)$ with fixed $\ellipsoidalSet_I, \ellipsoidalSet_T$, and $\mathrm{iter}_{\max} = 4$. Each cell entry: funnel volume \eqref{eq:approximate_funnel_volume} $|$ inlet-to-outlet volume ratio $\mathrm{vol}(\levelSet_0)/\mathrm{vol}(\levelSet_K)$. 
`--': the initial $(V, \rho)$ guess fails, i.e., L-step (line~\ref{algline:L_step}, Algorithm~\ref{alg:bilinear_alternation}) is infeasible. `C': the initial guess violates inlet or outlet containment. 
}
\label{tab:sweep_funnel_chapter}
\centering
\begin{tabular}{c|cccc}
\hline
$\rho_{\mathrm{g}} \downarrow \backslash c \rightarrow$ & 1 & 1.5 & 2 & 3 \\
\hline
0.02 & C & C & 0.152 $|$ 2355 & -- \\
0.035 & C & 0.159 $|$ 2407 & 0.153 $|$ 2779 & -- \\
0.05 & 0.155 $|$ 2205 & 0.157 $|$ 2400 & -- & -- \\
0.0625 & C & C & C & C \\
\hline
\end{tabular}
\end{table}

\subsection{Discussion}
\label{subsec:results_discussion_funnel_chapter}

\subsubsection{Alternation procedure}

Fig.~\ref{fig:alternation_funnel_chapter} shows objective function value \eqref{eq:sos_cost} and funnel volume over 20 iterations of Algorithm~\ref{alg:bilinear_alternation} on the default problem (Table~\ref{tab:setup_funnel_chapter}). The objective decreases monotonically, as guaranteed, but with a long tail: the stopping criterion (via convergence tolerance $\delta$) is met at $\mathrm{iter} = 4$, which already captures 89.9\% of the decrease obtained after 20 iterations. The funnel volume falls by 82.5\% from the initial guess to termination and then plateaus without decreasing monotonically: its minimum over 20 iterations ($\mathrm{iter} = 16$) is 4.4\% below the value at termination.
The funnel volume does not follow the objective in this tail; this decoupling is a practical consequence of objective \eqref{eq:sos_cost} being a volume proxy. Once the objective value's tail is reached ($\mathrm{iter} > 4$), further iterations reshape the funnel without reliably shrinking it, and an early termination of Algorithm~\ref{alg:bilinear_alternation} results in significant computation gains, albeit forgoing some optimality gains.

\subsubsection{Polynomial approximation validity}
\label{subsubsec:validity_funnel_chapter}
The SOS-based invariance certificate holds for the Taylor-approximated dynamics $\widetilde{\mathbf{f}}_k$. To test it against the true deviation dynamics $\bar{\mathbf{f}}_k$, we sample 1000 points on the boundary of each $\levelSet_k$ (directions sampled uniformly) and evaluate the sampled invariance condition \eqref{eq:Lyapunov_invariance_condition_discretized} with $\bar{\mathbf{f}}_k$ in place of $\widetilde{\mathbf{f}}_k$. Test funnels are computed with scaled inlets $s\,\ellipsoidalSet_I$ and correspondingly scaled initial guesses. For $s \in \{0.5, 1, 1.25\}$, none of the 39{,}000 boundary samples per funnel violated the condition. 
The worst-case normalized excess over all knot points $k$ and boundary samples $\bar{\stateVector}_s \in \partial\levelSet_k$, $\max_{k} \max_{\bar{\stateVector}_s} \big(\dot{V}_k(\bar{\stateVector}_s) - \dot{\rho}_k\big)/|\dot{\rho}_k|$, with $\dot{V}_k$ evaluated under $\bar{\mathbf{f}}_k$, is negative in all three funnels (i.e., satisfying \eqref{eq:Lyapunov_invariance_condition_discretized}, $V_k = \rho_k \implies \dot{V}_k - \dot{\rho}_k \le 0$), ranging from $-2.2\%$ to $-0.6\%$. 
These results are empirical evidence over the tested range, not a guarantee, and do not address behavior between knot points.

\subsubsection{Computation times}
Let $T(s) \triangleq s(s+1)/2$ denote the number of free entries of an $s \times s$ symmetric Gram matrix. With quadratic $V_k$ and Taylor degree $\deg\widetilde{\mathbf{f}}_k = 3$, the invariance constraint \eqref{eq:S_procedure_invariance_constraint} is a polynomial of degree
\begin{align}
\begin{split}
    2d &= \max\big(\deg V_k - 1 + \deg\widetilde{\mathbf{f}}_k,\; \deg\lambda_k + \deg V_k\big) \\ &= \max\big(4,\, \deg\lambda_k + 2\big),
\end{split}
    \label{eq:invariance_constraint_degree}
\end{align}
so invariance constraint at each knot point contributes one Gram block of size $g = \binom{n+d}{d}$. The blocks corresponding to remaining constraints are: one of size $n$ for each positive-definiteness constraint \eqref{eq:sos_pd}, a homogeneous quadratic; one of size $n+1$ for each containment constraint \eqref{eq:sos_initial}--\eqref{eq:sos_terminal}, a quadratic with a constant term; and $1 \times 1$ blocks for the scalar SOS variables. 
The number of scalarized PSD entries (SDP decision variables) corresponding to the invariance constraints at all knot points alone is: $K\,T(g) = K\binom{n+d}{d}\big(\binom{n+d}{d}+1\big)/2$, which grows combinatorially in $n$ and $d$ and dominates other PSD entries (decision variables).

\begin{table*}[t!]
\caption{Cost per alternation iteration of Algorithm~\ref{alg:bilinear_alternation}, averaged over three iterations. Gram size $g = \binom{n+d}{d}$, with $d$ given by \eqref{eq:invariance_constraint_degree}. PSD entries: scalarized entries of all psd blocks as formulated (L-step / V-step). Times: wall-clock seconds, with MOSEK solver time in parentheses. $^\dagger$Analytic count (the run did not complete on the test machine).}
\label{tab:timing_funnel_chapter}
\centering
\begin{tabular}{lcccccccc}
\hline
\rule{0pt}{2ex}System & $n$ & $K\!+\!1$ & $\deg\lambda_k$ & Gram size & PSD entries & L-step [s] & V-step [s] \\
\hline
\rule{0pt}{2ex}Unicycle & 3 & 40 & 2 & 10 & 2401 / 2446 & 3.11 (0.029) & 4.73 (0.078) \\
Unicycle & 3 & 40 & 4 & 20 & 8446 / 8491 & 6.43 (0.078) & 9.11 (0.21) \\
Unicycle & 3 & 40 & 6 & 35 & 24826 / 24871 & 13.3 (0.26) & 17.7 (0.61) \\
Cart-pole & 4 & 25 & 2 & 15 & 3152 / 3186 & 2.87 (0.039) & 4.19 (0.11) \\
Cart-pole & 4 & 25 & 4 & 35 & 15392 / 15426 & 8.72 (0.26) & 12.1 (0.6) \\
Cart-pole & 4 & 25 & 6 & 70 & 59912 / 59946 & 29.5 (0.96) & 36.3 (2.3) \\
Quadrotor & 12 & 25 & 2 & 91 & 102520 / 102622 & 119.3 (10.8) & 157.3 (32.0) \\
Quadrotor & 12 & 25 & 4 & 455 & 2491816 / 2491918$^\dagger$ & -- & -- \\
\hline
\end{tabular}
\end{table*} 

Table~\ref{tab:timing_funnel_chapter} reports the cost per alternation iteration, averaged over three iterations, for the unicycle ($n=3$), a cart-pole ($n=4$), and a quadrotor ($n=12$). 
We observe that compute time is dominated by SOSTOOLS constructing the SDP rather than in solving it: MOSEK accounts for at most 6.4\% of the wall time for $n \le 4$, rising to 9--20\% for the quadrotor. Since the structure of both SDPs (L-step \& V-step) is identical across iterations and only their coefficients change, this overhead could be paid once per problem rather than once per step.

Raising $\deg\lambda_k$ from 2 to 6 multiplies the wall time by $4\times$ (unicycle) and $9\times$ (cart-pole), but shrinks the funnel volume after three iterations by only 3.8\% and 9.4\%, respectively. At $\deg\lambda_k = 2$, moving from unicycle to quadrotor increases the PSD entries per knot by about $70\times$, wall time per knot by $57\times$, and MOSEK time per knot by $650\times$ --- solver's share of the compute cost quickly grows with $n$, and a single iteration (L-step + V-step) for the quadrotor system takes $\sim\,$4.6 minutes. At $\deg\lambda_k = 4$, the quadrotor SDP grows to $2.5 \times 10^6$ PSD entries, $24$ times the degree-2 problem, and the run did not complete on our hardware. %

\begin{figure}[t!]
\centering
\includegraphics[width=0.8\columnwidth]{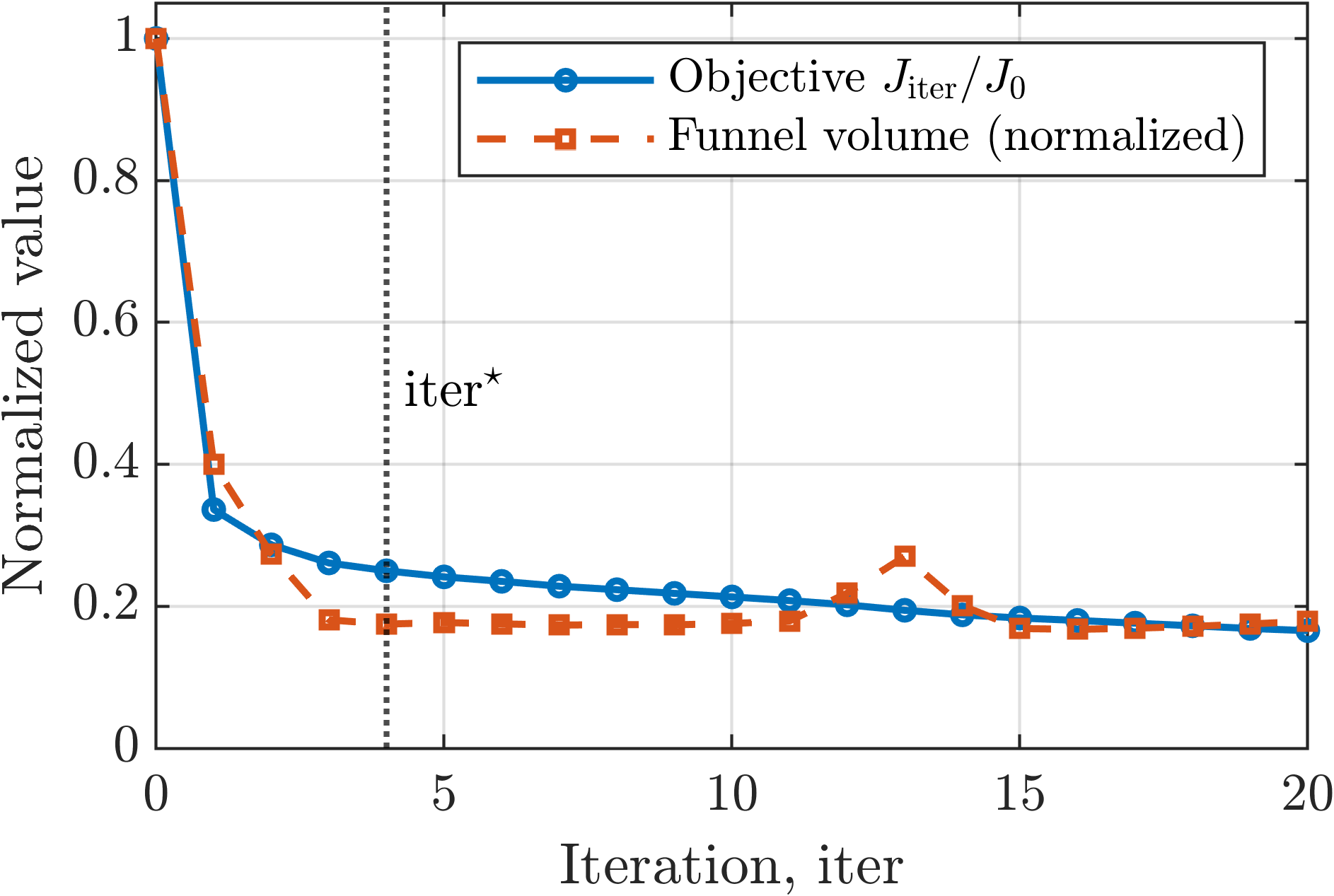}
\caption{Objective $J_{\mathrm{iter}}$ and funnel volume \eqref{eq:approximate_funnel_volume} over 20 iterations of Algorithm~\ref{alg:bilinear_alternation} on the default problem, each normalized by its value at the initial guess ($\mathrm{iter} = 0$). The dotted line at $\mathrm{iter} = 4$ marks termination under convergence tolerance $\delta = 0.05$. The objective $J$ is monotone by construction (the funnel volume need not be).}
\label{fig:alternation_funnel_chapter}
\end{figure}

\subsubsection{Limitations}

Funnel invariance certificate obtained using our proposed formulation is approximate in three ways. 
First, it only holds for the Taylor-approximated dynamics: truncation error is bounded only locally, and is largest on the funnel boundary where invariance is checked. We empirically found no violations over the certified funnel (Section~\ref{subsec:results_discussion_funnel_chapter}), but a guarantee requires bounding the remainder term (truncation error) over the funnel and treating it as a bounded disturbance in the SOS program. 
Second, invariance is certified at discrete knot points only, with no inter-sample guarantee. 
Third, collocation residual of the nominal trajectory is neglected.

Beyond the shortcomings of discrete-time numerical SOS analysis, bilinear alternation solves a non-convex problem via sequential convex steps: it converges in objective value only, offers no guarantee of global optimality, and its result depends on the warm start (Table~\ref{tab:sweep_funnel_chapter}). Moreover, the proxy objective \eqref{eq:sos_cost} may not track funnel volume monotonically (Fig.~\ref{fig:alternation_funnel_chapter}), which a log-determinant volume objective would address \citep{vandenberghe1998determinant}.
Finally, the Gram matrix dimension grows combinatorially in state dimensionality $n$ --- DSOS/SDSOS relaxations \citep{ahmadi2019dsos, majumdar2020recent} could improve scalability at the cost of conservatism.

\section{Summary}
\label{sec:summary_funnel_chapter}

This paper presented an end-to-end pipeline for computing invariant funnels with SOS programming, its open-source implementation \texttt{computeSOSFunnels}, and an empirical characterization of its behavior in practice. Bounding the outlet by a terminal set nested inside the inlet set makes funnels with coinciding nominal endpoints composable by construction. Numerical experiments show that infeasible initial guesses are partly decidable before solving, the alternation procedure can be stopped after a few iterations with little loss in funnel volume, and the certificate computed for the Taylor-approximated dynamics held under the true dynamics at every tested boundary sample. 
Low polynomial multiplier degrees are the economical choice: raising the degree from 2 to 6 costs $4$--$9\times$ the compute wall-time for only a 4--10\% reduction in funnel volume. Most of the computation time is spent constructing the SDP rather than solving it. Future work will bound the Taylor remainder to certify the true dynamics, adopt a log-determinant volume objective, and reuse the SDP structure across iterations to reduce the computational overhead.


\section*{Acknowledgments}
The author thanks his advisor, Prof. Michael Otte, for his guidance throughout this work, and is extremely grateful to Sandeep Kumar, Abraham Vinod, and Purnanand Elango for the insightful discussions on numerical optimization.

\bibliographystyle{IEEEtranN}


\begin{thebibliography}{26}
\providecommand{\natexlab}[1]{#1}
\providecommand{\url}[1]{#1}
\csname url@samestyle\endcsname
\providecommand{\newblock}{\relax}
\providecommand{\bibinfo}[2]{#2}
\providecommand{\BIBentrySTDinterwordspacing}{\spaceskip=0pt\relax}
\providecommand{\BIBentryALTinterwordstretchfactor}{4}
\providecommand{\BIBentryALTinterwordspacing}{\spaceskip=\fontdimen2\font plus
\BIBentryALTinterwordstretchfactor\fontdimen3\font minus \fontdimen4\font\relax}
\providecommand{\BIBforeignlanguage}[2]{{%
\expandafter\ifx\csname l@#1\endcsname\relax
\typeout{** WARNING: IEEEtranN.bst: No hyphenation pattern has been}%
\typeout{** loaded for the language `#1'. Using the pattern for}%
\typeout{** the default language instead.}%
\else
\language=\csname l@#1\endcsname
\fi
#2}}
\providecommand{\BIBdecl}{\relax}
\BIBdecl

\bibitem[Burridge et~al.(1999)Burridge, Rizzi, and Koditschek]{burridge1999sequential}
R.~R. Burridge, A.~A. Rizzi, and D.~E. Koditschek, ``Sequential composition of dynamically dexterous robot behaviors,'' \emph{The International Journal of Robotics Research}, vol.~18, no.~6, pp. 534--555, 1999.

\bibitem[Tedrake et~al.(2010)Tedrake, Manchester, Tobenkin, and Roberts]{Tedrake.etal.IJRR10}
R.~Tedrake, I.~R. Manchester, M.~Tobenkin, and J.~W. Roberts, ``{LQR}-trees: Feedback motion planning via sums-of-squares verification,'' \emph{The International Journal of Robotics Research}, vol.~29, no.~8, pp. 1038--1052, 2010.

\bibitem[Majumdar and Tedrake(2017)]{Majumdar.Tedrake.IJRR17}
A.~Majumdar and R.~Tedrake, ``Funnel libraries for real-time robust feedback motion planning,'' \emph{The International Journal of Robotics Research}, vol.~36, no.~8, pp. 947--982, 2017.

\bibitem[Jaffar and Otte(2023)]{Jaffar.Otte.WAFR22}
M.~K.~M. Jaffar and M.~Otte, ``{PiP-X}: Funnel-based online feedback motion planning/replanning in dynamic environments,'' in \emph{International Workshop on the Algorithmic Foundations of Robotics}.\hskip 1em plus 0.5em minus 0.4em\relax Springer, 2023, pp. 132--148.

\bibitem[Parrilo(2000)]{parrilo2000structured}
P.~A. Parrilo, ``Structured semidefinite programs and semialgebraic geometry methods in robustness and optimization,'' Ph.D. dissertation, California Institute of Technology, 2000.

\bibitem[Papachristodoulou and Prajna(2002)]{papachristodoulou2002construction}
A.~Papachristodoulou and S.~Prajna, ``On the construction of {Lyapunov} functions using the sum of squares decomposition,'' in \emph{Proceedings of the 41st IEEE Conference on Decision and Control}, vol.~3, 2002, pp. 3482--3487.

\bibitem[Tobenkin et~al.(2011)Tobenkin, Manchester, and Tedrake]{Tobenkin.etal.IFAC11}
M.~M. Tobenkin, I.~R. Manchester, and R.~Tedrake, ``Invariant funnels around trajectories using sum-of-squares programming,'' \emph{IFAC Proceedings Volumes}, vol.~44, no.~1, pp. 9218--9223, 2011.

\bibitem[Majumdar et~al.(2013)Majumdar, Ahmadi, and Tedrake]{Majumdar.Ahmadi.Tedrake.ICRA13}
A.~Majumdar, A.~A. Ahmadi, and R.~Tedrake, ``Control design along trajectories with sums of squares programming,'' in \emph{Proceedings of the IEEE International Conference on Robotics and Automation}, 2013, pp. 4054--4061.

\bibitem[Jaffar and Otte(2022)]{jaffar2022pip}
M.~K.~M. Jaffar and M.~Otte, ``{PiP-X: Online feedback motion planning/replanning in dynamic environments using invariant funnels},'' \emph{arXiv preprint arXiv:2202.00772}, 2022.

\bibitem[Kim et~al.(2023)Kim, Elango, Reynolds, A{\c{c}}{\i}kme{\c{s}}e, and Mesbahi]{Kim2023}
T.~Kim, P.~Elango, T.~P. Reynolds, B.~A{\c{c}}{\i}kme{\c{s}}e, and M.~Mesbahi, ``Optimization-based constrained funnel synthesis for systems with {Lipschitz} nonlinearities via numerical optimal control,'' \emph{IEEE Control Systems Letters}, vol.~7, pp. 2875--2880, 2023.

\bibitem[Herbert et~al.(2017)Herbert, Chen, Han, Bansal, Fisac, and Tomlin]{Herbert.etal.CDC17}
S.~L. Herbert, M.~Chen, S.~Han, S.~Bansal, J.~F. Fisac, and C.~J. Tomlin, ``{FaSTrack}: A modular framework for fast and guaranteed safe motion planning,'' in \emph{Proceedings of the IEEE Conference on Decision and Control}, 2017, pp. 1517--1522.

\bibitem[Seo et~al.(2021)Seo, Son, and Kim]{Seo2021}
H.~Seo, C.~Y. Son, and H.~J. Kim, ``Fast funnel computation using multivariate {Bernstein} polynomial,'' \emph{IEEE Robotics and Automation Letters}, vol.~6, no.~2, pp. 1351--1358, 2021.

\bibitem[Singh et~al.(2023)Singh, Landry, Majumdar, Slotine, and Pavone]{Singh2023}
S.~Singh, B.~Landry, A.~Majumdar, J.-J. Slotine, and M.~Pavone, ``Robust feedback motion planning via contraction theory,'' \emph{The International Journal of Robotics Research}, vol.~42, no.~9, pp. 655--688, 2023.

\bibitem[Ahmadi and Majumdar(2019)]{ahmadi2019dsos}
A.~A. Ahmadi and A.~Majumdar, ``{DSOS} and {SDSOS} optimization: More tractable alternatives to sum of squares and semidefinite optimization,'' \emph{SIAM Journal on Applied Algebra and Geometry}, vol.~3, no.~2, pp. 193--230, 2019.

\bibitem[Majumdar et~al.(2020)Majumdar, Hall, and Ahmadi]{majumdar2020recent}
A.~Majumdar, G.~Hall, and A.~A. Ahmadi, ``Recent scalability improvements for semidefinite programming with applications in machine learning, control, and robotics,'' \emph{Annual Review of Control, Robotics, and Autonomous Systems}, vol.~3, pp. 331--360, 2020.

\bibitem[Blekherman et~al.(2012)Blekherman, Parrilo, and Thomas]{blekherman2012semidefinite}
G.~Blekherman, P.~A. Parrilo, and R.~R. Thomas, Eds., \emph{Semidefinite Optimization and Convex Algebraic Geometry}, ser. MOS-SIAM Series on Optimization.\hskip 1em plus 0.5em minus 0.4em\relax SIAM, 2012, vol.~13.

\bibitem[Papachristodoulou and Prajna(2005)]{papachristodoulou2005analysis}
A.~Papachristodoulou and S.~Prajna, ``Analysis of non-polynomial systems using the sum of squares decomposition,'' in \emph{Positive Polynomials in Control}, ser. Lecture Notes in Control and Information Sciences.\hskip 1em plus 0.5em minus 0.4em\relax Springer, 2005, vol. 312, pp. 23--43.

\bibitem[Parrilo(2003)]{parrilo2003semidefinite}
P.~A. Parrilo, ``Semidefinite programming relaxations for semialgebraic problems,'' \emph{Mathematical programming}, vol.~96, no.~2, pp. 293--320, 2003.

\bibitem[Boyd et~al.(1994)Boyd, El~Ghaoui, Feron, and Balakrishnan]{boyd1994linear}
S.~Boyd, L.~El~Ghaoui, E.~Feron, and V.~Balakrishnan, \emph{Linear Matrix Inequalities in System and Control Theory}, ser. Studies in Applied Mathematics.\hskip 1em plus 0.5em minus 0.4em\relax SIAM, 1994, vol.~15.

\bibitem[Nagumo(1942)]{nagumo1942lage}
M.~Nagumo, ``{\"U}ber die {L}age der {I}ntegralkurven gew{\"o}hnlicher {D}ifferentialgleichungen,'' \emph{Proceedings of the Physico-Mathematical Society of Japan, 3rd Series}, vol.~24, pp. 551--559, 1942.

\bibitem[Blanchini(1999)]{blanchini1999set}
F.~Blanchini, ``Set invariance in control,'' \emph{Automatica}, vol.~35, no.~11, pp. 1747--1767, 1999.

\bibitem[Kelly(2017)]{kelly2017introduction}
M.~Kelly, ``An introduction to trajectory optimization: How to do your own direct collocation,'' \emph{SIAM Review}, vol.~59, no.~4, pp. 849--904, 2017.

\bibitem[Bryson and Ho(2018)]{bryson2018applied}
A.~E. Bryson and Y.-C. Ho, \emph{Applied Optimal Control: Optimization, Estimation, and Control}.\hskip 1em plus 0.5em minus 0.4em\relax Routledge, 2018.

\bibitem[Papachristodoulou et~al.(2021)Papachristodoulou, Anderson, Valmorbida, Prajna, Seiler, Parrilo, Peet, and Jagt]{SOSTOOLS}
A.~Papachristodoulou, J.~Anderson, G.~Valmorbida, S.~Prajna, P.~Seiler, P.~A. Parrilo, M.~M. Peet, and D.~Jagt, \emph{{SOSTOOLS}: Sum of Squares Optimization Toolbox for {MATLAB}}, 2021, version 4.00. Available from \url{https://github.com/oxfordcontrol/SOSTOOLS}; arXiv:1310.4716.

\bibitem[{MOSEK ApS}(2024)]{MOSEK}
\BIBentryALTinterwordspacing
{MOSEK ApS}, \emph{The {MOSEK} Optimization Toolbox for {MATLAB} Manual. Version 10.2}, 2024. [Online]. Available: \url{https://docs.mosek.com/10.2/toolbox/index.html}
\BIBentrySTDinterwordspacing

\bibitem[Vandenberghe et~al.(1998)Vandenberghe, Boyd, and Wu]{vandenberghe1998determinant}
L.~Vandenberghe, S.~Boyd, and S.~P. Wu, ``Determinant maximization with linear matrix inequality constraints,'' \emph{SIAM Journal on Matrix Analysis and Applications}, vol.~19, no.~2, pp. 499--533, 1998.

\end{thebibliography}

\end{document}